\documentclass[9pt,twocolumn,twoside]{pnas-new}

\templatetype{pnasresearcharticle} 

\title{Detecting abnormality in heart dynamics from multifractal analysis of ECG signals}

\author[a]{Snehal M. Shekatkar}
\author[a]{Yamini Kotriwar} 
\author[b]{K.P. Harikrishnan}
\author[a,1]{G. Ambika}

\affil[a]{Indian Institute of Science Education and Research, Pune-411008, India}
\affil[b]{Department of Physics, The Cochin College, Cochin-682002, India}

\leadauthor{Shekatkar} 

\significancestatement{We address an important topic in clinical sciences about extracting quantifiers from measured Electro Cardio Gram(ECG)data to distinguish normal heart rhythms from the abnormal ones. The computed multifractal measures derived from discretized ECG, combined with a machine learning approach, provide fairly accurate indications of abnormality in cardiac rhythms. Further the subtle variations in the shape of beats captured from the analysis of beat replicated data have relevance in detecting variations in the normal heart dynamics. The analysis can enhance our understanding of possible transitions in the nonlinear dynamics underlying ECG data.}

\authorcontributions{G.A. and K.P.H. conceived the research. Y.K. downloaded and detrended the data. S.M.S. automated the codes, performed the analysis. G.A. and S.M.S. prepared the manuscript. All authors discussed and reviewed the manuscript.}
\authordeclaration{Authors declare no conflict of interest.}
\correspondingauthor{\textsuperscript{1}To whom correspondence should be addressed. E-mail: g.ambika\@iiserpune.ac.in}


\begin{abstract}
The characterization of heart dynamics with a view to distinguish abnormal from normal behavior is an interesting topic in clinical sciences. Here we present an analysis of the Electro-cardiogram (ECG) signals obtained under controlled conditions from several healthy and unhealthy subjects using the framework of multifractal analysis. Our analysis differs from the conventional nonlinear analysis in that the information contained in the amplitude variations of the signal is being extracted and quantified. The results thus obtained reveal that the attractor underlying the dynamics of the heart has multifractal structure and the resultant multifractal spectra can clearly separate healthy subjects from unhealthy ones. We use supervised machine learning approach to build a model that predicts the group label of a new subject with very high accuracy on the basis of the multifractal parameters. By comparing the range of scaling indices in the multifractal spectra with that of beat replicated data from the same ECG, we show how each ECG can be checked for abnormality for variations within itself. 
\end{abstract}

\dates{This manuscript was compiled on \today}
\doi{\url{www.pnas.org/cgi/doi/10.1073/pnas.XXXXXXXXXX}}

\begin{document}

\verticaladjustment{-2pt}

\maketitle
\thispagestyle{firststyle}
\ifthenelse{\boolean{shortarticle}}{\ifthenelse{\boolean{singlecolumn}}{\abscontentformatted}{\abscontent}}{}


\dropcap{T}he complexity of many physiological rhythms originate from the underlying complex nonlinear dynamical processes \cite{lewis1975phase, glass2001synchronization, karma2013physics, pijn1991chaos, qu2014nonlinear, webber1994dynamical, muller2016causality}. The various levels of this complexity and their variations, if properly discerned, will be useful in understanding the abnormalities that lead to many pathological cases\cite{mendis2011global, world2014global, go2014heart}. A mathematical representation of their complexity using dynamical equations is most often not realizable due to the many interacting variables and uncertain parameters involved \cite{bruggeman2007nature, lillacci2010parameter}. Hence the only possibility to study the dynamics in such situations to is to rely on information that can be deciphered from signals obtained from these systems\cite{kantz2004nonlinear, lehnertz1998can, richman2000physiological}. Over the last few decades, several such physiological signals like EEG, ECG, fMRI etc have been subjected to different techniques available under the broad area of nonlinear time series analysis \cite{kurths1995quantitative, marwan2002recurrence, kantz2004nonlinear}.

It is established that most of the cardiovascular diseases arise from the changes in the dynamics of the heart. A healthy heart is a complex system with fractal nature \cite{ivanov1999multifractality, yang2007multifractal, glass2012theory} but cardiac abnormalities or malfunctions can cause subtle changes or variations in its complexity. However, the complexity related quantifiers have not yet reached the clinics for effective diagnostics and therapy. For this, we have to develop a unique way of characterizing its complexity that will help to distinguish healthy and pathological cases. In the present work, we report how measures derived from the multi fractal spectrum of ECG signals can effectively be used as a promising tool in the diagnosis of abnormalities of the heart. 

There are a few reported research in related analysis using ECG signals \cite{kurths1995quantitative, voss1996application, ivanov1999multifractality, yang2007multifractal}. But most of them are on the peak to peak time intervals (also called as \textit{R-R} intervals) \cite{makowiec2011reading, cysarz2011unexpected} and hence do not include the possible information content in the amplitude variations in the signals. This motivates us to proceed with a detailed analysis that provides unique measures that characterize the actual amplitude variations also from the point of view of the multifractal analysis. Such analysis is also advantageous over \textit{R-R} interval analysis because the actual amount of time for which the signal needs to be recorded is smaller by orders of magnitude.

As already mentioned, dynamics underlying many natural processes are essentially deterministic but are highly nonlinear and complex. Since generally the dynamical equations or even the effective dimension of the system are not known, we have to rely on observational data of one of the variables or time series of the average responses like ECG to characterize its dynamics. The first step in the analysis is the reconstruction of the underlying attractor from the time series of a single variable \cite{kantz2004nonlinear} based on Taken's theorem \cite{takens1981dynamical, packard1980geometry}. Over the last few decades, this has become an extremely mature field and a wide variety of methods have been developed \cite{takens1981dynamical, packard1980geometry, theiler1992testing, schreiber2000surrogate, grassberger1983measuring, harikrishnan2006non, harikrishnan2009computing}. These methods have found successful applications in diverse fields like astrophysics \cite{misra2006nonlinear}, physiology \cite{voss1996application, perc2005nonlinear}, atmospheric sciences \cite{vautard1989singular, koccak2000nonlinear}, geology \cite{foufoula2014wavelets} and stock markets \cite{hafner2013nonlinear}. Among them, an important class of methods is related to the characterization of the complex dynamics using multifractal analysis \cite{kantz2004nonlinear, mandelbrot2013multifractals}.

We identify two such measures computed from the multi fractal spectrum with their ranges distinctly distinguishable for healthy and diseased cases. This makes the proposed analysis very effective and powerful for clinical purposes where better accuracy compared to visual inspection can be realized. We illustrate this by applying the analysis to ECG signals from a number of normal and pathological subjects and our conclusions are validated using supervised machine learning approach and comparisons with beat replicated data from individual ECG.

\section{Phase space reconstruction from ECG data}
\begin{figure}[h]
\begin{center}
\includegraphics[trim=50 0 100 0, clip=true,width=\columnwidth]{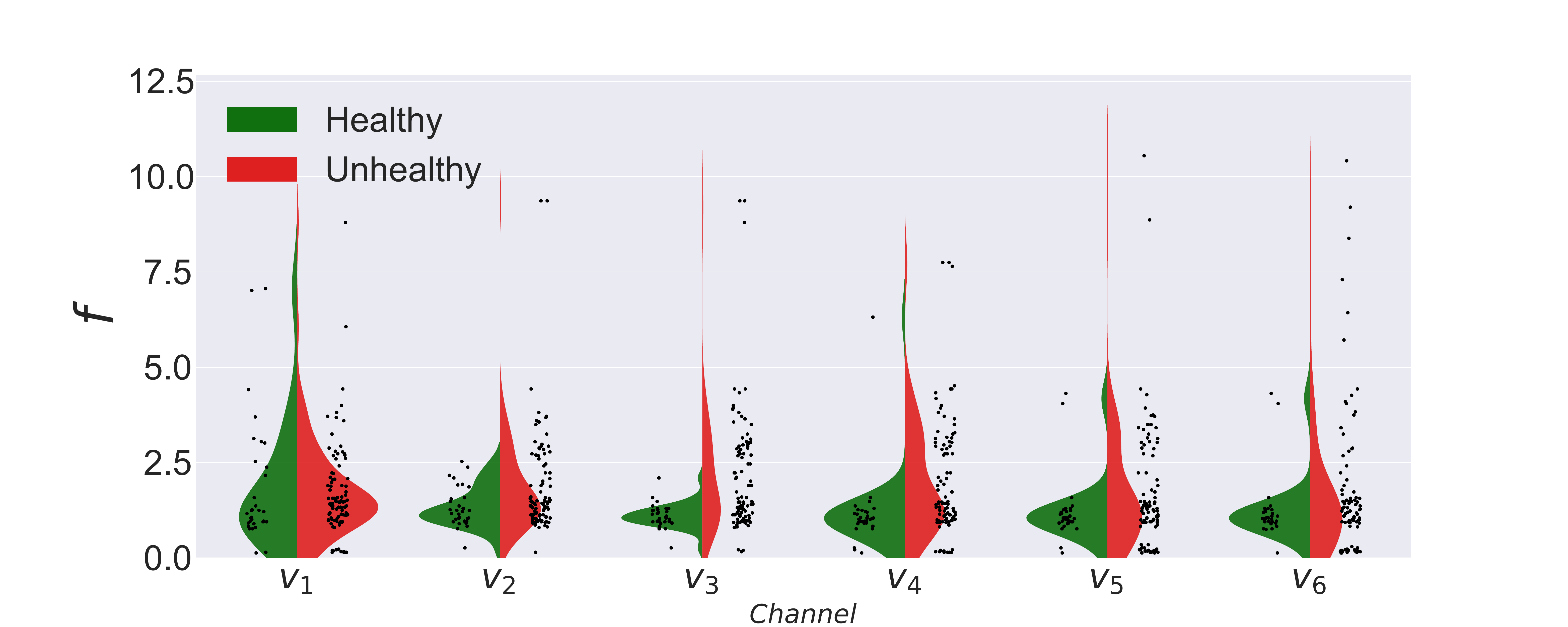}
\caption{\label{peak_freq} Violin-plot showing the distributions of frequencies with highest power for healthy and unhealthy groups for different chest electrodes. }
\end{center}
\end{figure}

The data of $97$ unhealthy and $32$ healthy subjects obtained from PhysioBank database \cite{goldberger2000physiobank} are pre-processed to make them suitable for the analysis (See Methods). Each dataset consists of ECG time series taken from six different chest electrodes or channels $v_1$ to $v_6$. As a preliminary analysis, we carry out the usual statistical and linear analysis and obtain the power spectra for all the data sets using the Fast Fourier Transform (FFT) algorithm. The frequency with the maximum power for each one of them and the distribution of these peak frequencies for all the subjects are summarized in Fig.~\ref{peak_freq} for data from six chest electrodes. It is clear that it is not possible to conclusively distinguish the two groups using the peak frequencies as they fall in similar ranges. 

\begin{figure}[h]
\begin{center}
\includegraphics[width=0.95\columnwidth, trim = 20 20 20 20, clip = true]{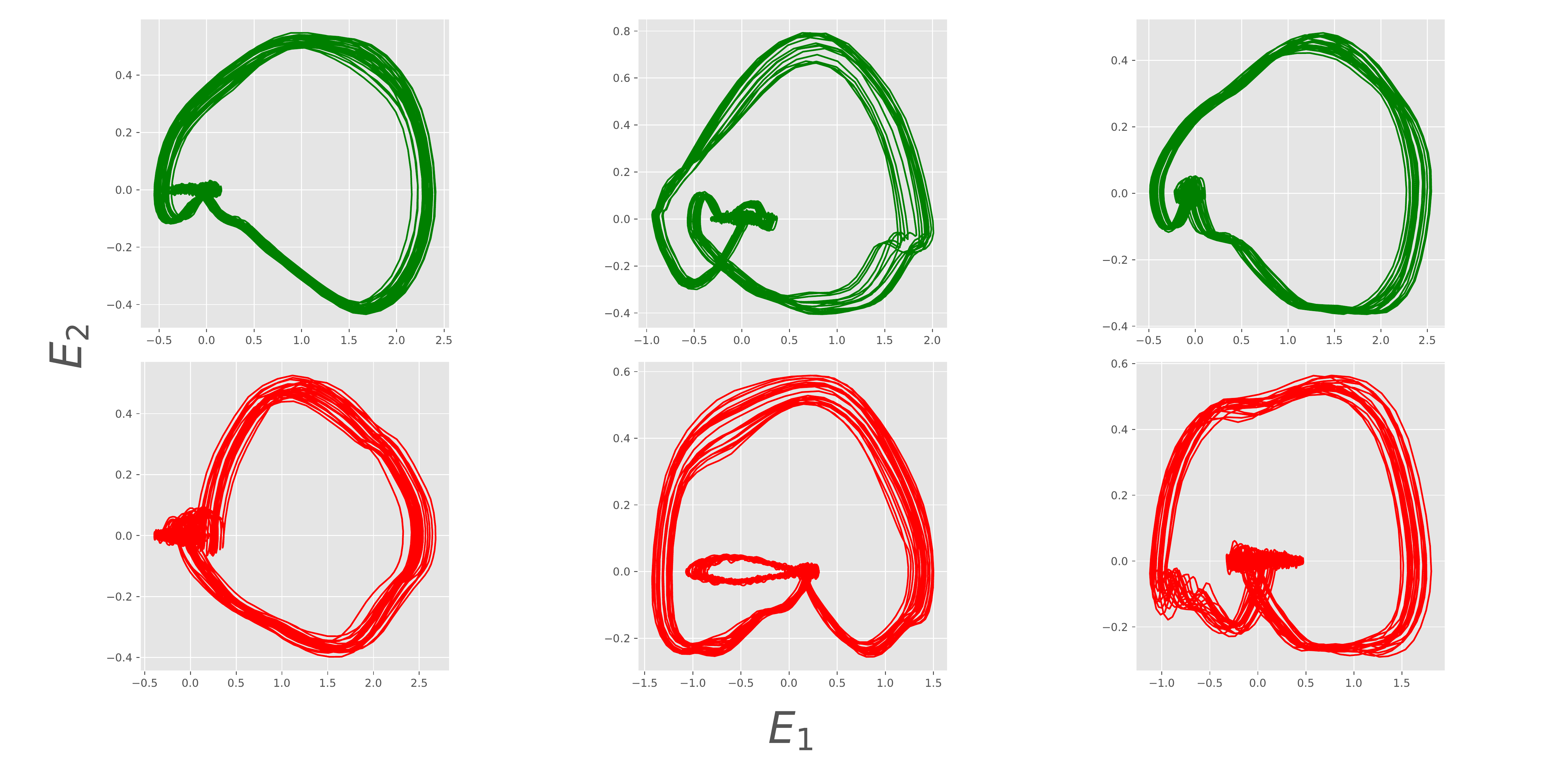}
\caption{\label{phase_space} Two dimensional projections of embedded attractors for ECG signals for three healthy (top panel) and three unhealthy (bottom panel) cases. All the attractors are plotted using the SVD transformation.}
\end{center}
\end{figure}

Hence we resort to methods of nonlinear analysis and reconstruct phase space structure of the system's dynamics from its discretely sampled time series $s(t_k)$. For a visual display of the resulting phase space structure or dynamical attractor, we use the technique of singular-value decomposition (SVD) \cite{broomhead1986extracting, albano1988singular}. (See Methods section for details about embedding). In Fig.~\ref{phase_space} we show a few representative embedded attractors from healthy and unhealthy groups. Each of the attractors is shown in the axes corresponding to statistically independent variables obtained from SVD. 

Since the complex structure of the attractor inherently contains information about the complexity of its dynamics, it is important to characterize it quantitatively. Several quantifiers have been studied in this context over the years \cite{paladin1986intermittency}, the most effective among them being the set of fractal dimensions. In the following sections, we indicate how correlation dimension is used along with surrogate analysis to identify the nonlinear nature of the underlying dynamics and how the geometrical complexity of its structure is quantified using multifractal measures.

\section{Correlation dimension and Surrogate analysis}
\begin{figure}[h]
\includegraphics[trim = 30 0 0 0, clip = true, width=\columnwidth]{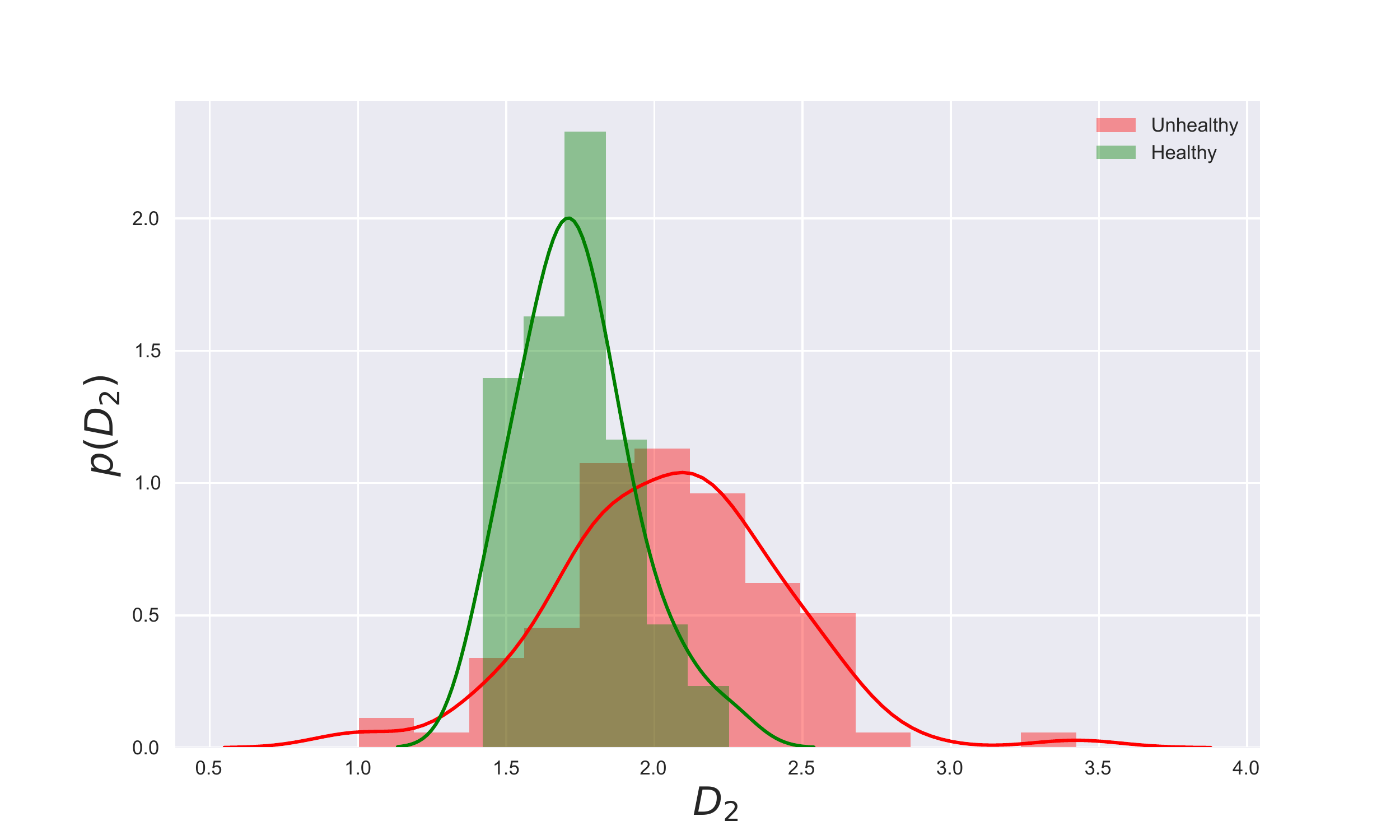}
\caption{\label{D2} Distribution of correlation dimension ($D_2$) values for healthy subjects (Green) and unhealthy ones (Red). As can be seen, all the values are less than $4$ and so we choose $M = 4$ as the embedding dimension.}
\end{figure}

We recreate the phase space structure of the underlying dynamics in an embedding space of dimension $M$ with delay vectors constructed by splitting the discretely sampled ECG data with delay time $\tau$. The dimension $M$ is chosen as the value at which any fractal measure, like correlation dimension $D_2$, saturates. The distributions of saturated $D_2$ values for healthy and unhealthy groups is shown in Fig.~\ref{D2}. Since all the $D_2$ values are less than $4$, for uniformity, we use $M = 4$ as the embedding dimension for all the signals. 

\begin{figure}[h]
\begin{center}
\includegraphics[width=0.8\columnwidth]{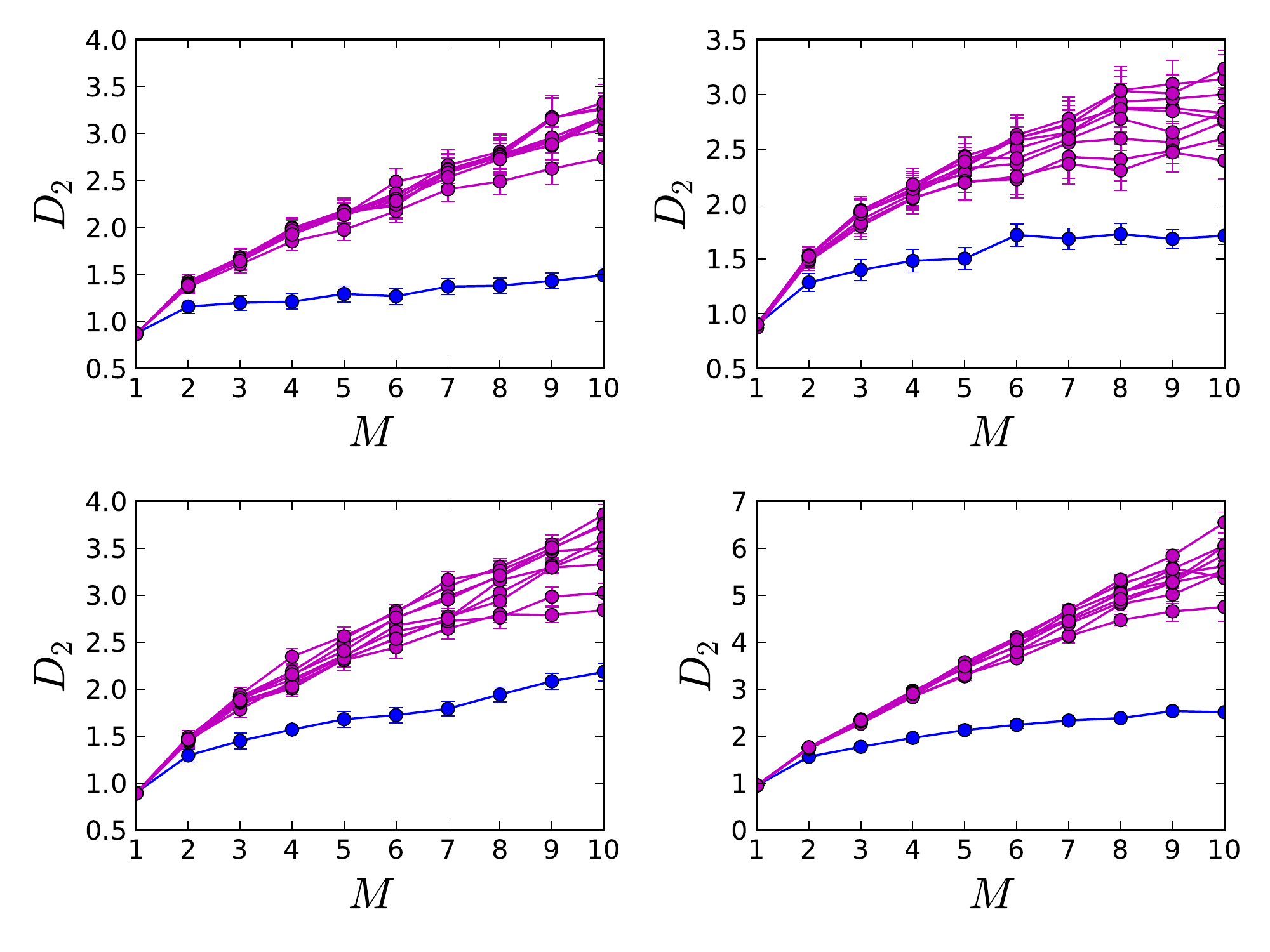}
\caption{\label{D2_saturate}  $D_2$ values as a function of embedding dimension $M$ for four randomly chosen datasets (blue) and their surrogates (magenta). It can be seen that the $D_2$ values of surrogates sufficiently deviate from that of the original data.}
\end{center}
\end{figure}

Before undertaking any type of nonlinear analysis, it is important to first verify that the observed time series indeed results from an underlying nonlinear process. We check the nonlinear and deterministic nature of the underlying dynamics using a statistically rigorous method of surrogate analysis\cite{theiler1992testing, schreiber2000surrogate}. For this, we generate surrogate data using TISEAN package \cite{hegger1999practical}. With correlation dimension as a measure, we plot in Fig.~\ref{D2_saturate} the values for the original signal and the generated surrogates as a function of embedding dimension $M$. Since the values of $D_2$ for surrogate data differ significantly from that of the original signal, we conclude that the signal comes from the underlying nonlinearity.

\section{\label{multifractal_analysis}Multifractal analysis}
The reconstructed phase-space attractors for most of the complex systems like heart dynamics possess a multifractal structure which is characterized by a set of generalized dimensions $D_q$ so that the non-uniformity in the distribution of points on the attractor becomes evident through different values of $q$. However the local scaling properties on the attractor are captured by a spectrum of singularities related to the probability measure on the course-grained attractor. Thus, if the attractor is covered by boxes of size $r$, the probability of points in the $i^{\text{th}}$ box scales as $p_i(r) \sim r^{\alpha_i}$. For a multifractal, the range of scales $\alpha_i$ present is a measure of its complexity. The number of boxes with the same $\alpha$ scales as $N_{\alpha}(r) \sim r^{f(\alpha)}$. Both $(D_q, q)$ and $(f(\alpha), \alpha)$ provide analogous characterizations and are related by Legendre transformations as \cite{harikrishnan2009computing}:

\begin{equation}
\label{Dq_alpha1}
\alpha = \frac{d}{dq}[(q-1)D_q]
\end{equation}

\begin{equation}
\label{Dq_alpha2}
f(\alpha) = q\alpha - (q-1)D_q
\end{equation}

Thus, a convenient way of calculating the multifractal spectrum is to calculate the generalized dimensions $D_q$ and then use equations (\ref{Dq_alpha1}) and (\ref{Dq_alpha2}) to find out $f(\alpha)$ and $\alpha$. An algorithmic approach to perform this has been given by Harikrishnan et al \cite{harikrishnan2009computing} and here we follow the same method with suitable modifications for ECG signals with the following mathematical form for $f(\alpha)$ spectrum:

\begin{equation}
f(\alpha) = A(\alpha-\alpha_1)^{\gamma_1}(\alpha_2-\alpha)^{\gamma_2}
\end{equation}
in which, as described in \cite{harikrishnan2009computing}, only four of the five parameters $A, \alpha_1, \alpha_2, \gamma_1, \gamma_2$ are independent. These four parameters provide a unique characterization of $f(\alpha)$ spectrum of the multifractal.

In our numerical computations of the four parameters from the multifractal analysis of ECG data, we find that the errors in the values of $\alpha_{2}$ and $\gamma_2$ are considerable. This is because they are derived from $f(\alpha)$ curve obtained from the $D_q$ data with negative values of $q$ which in turn corresponds to sparse regions of the attractor. With only a finite length of data, the number of vectors in the sparse regions of the attractor tends to become too low in numbers resulting in larger error bars for the $D_q$ values. For this reason, while deriving conclusions from the results, we concentrate mostly on the $\alpha_1$ and $\gamma_1$ values that are associated with the dense regions of the attractor.


\begin{figure}[h]
\begin{center}
\includegraphics[width=0.8\columnwidth]{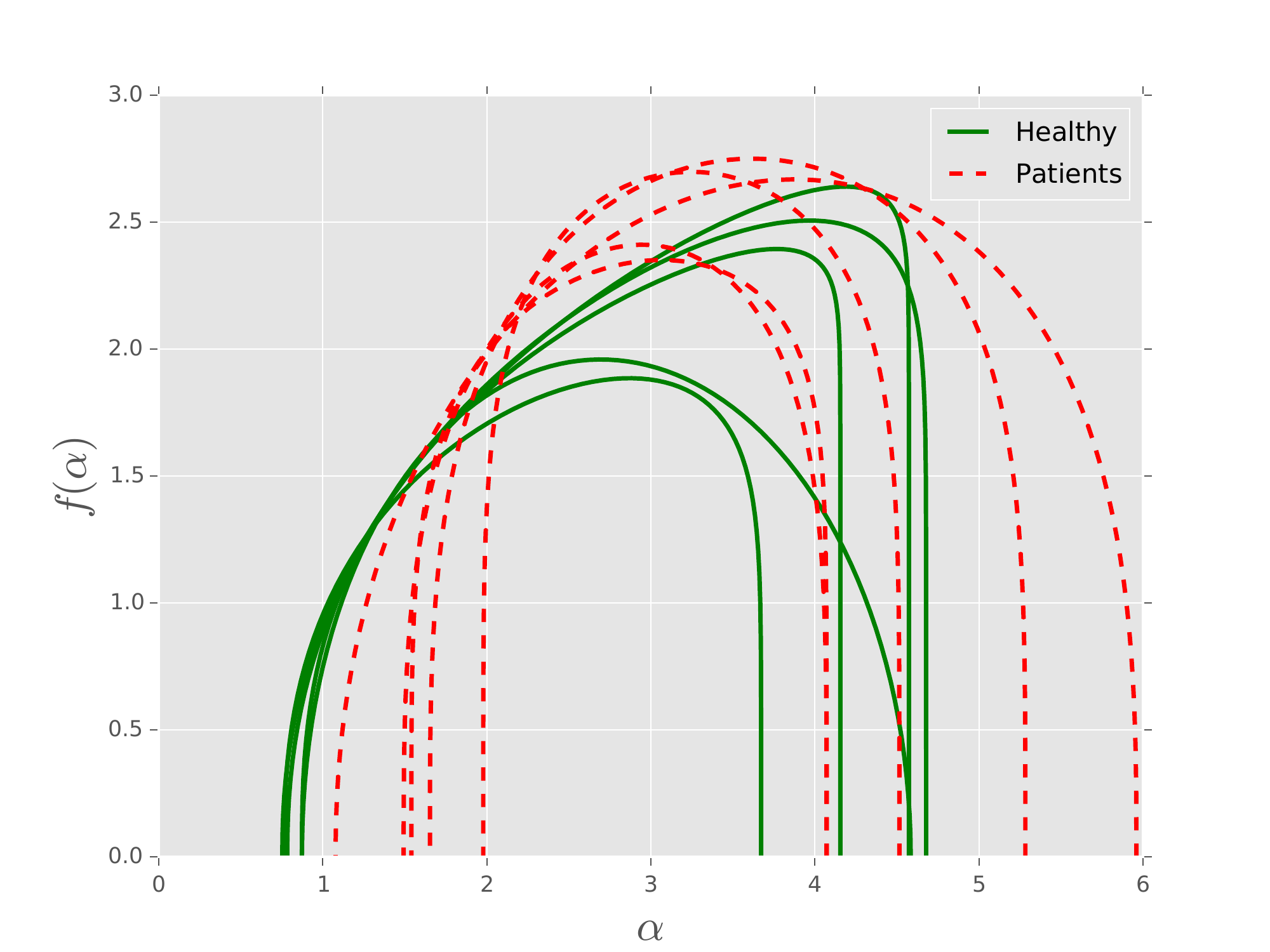}
\caption{\label{falpha} Multifractal spectrum curves for $5$ randomly selected subjects from each group. The continuous green curves represent healthy subjects while the dashed red curves are for the non-healthy ones. }
\end{center}
\end{figure}

\begin{figure}[h]
\begin{center}
\includegraphics[width=0.8\columnwidth]{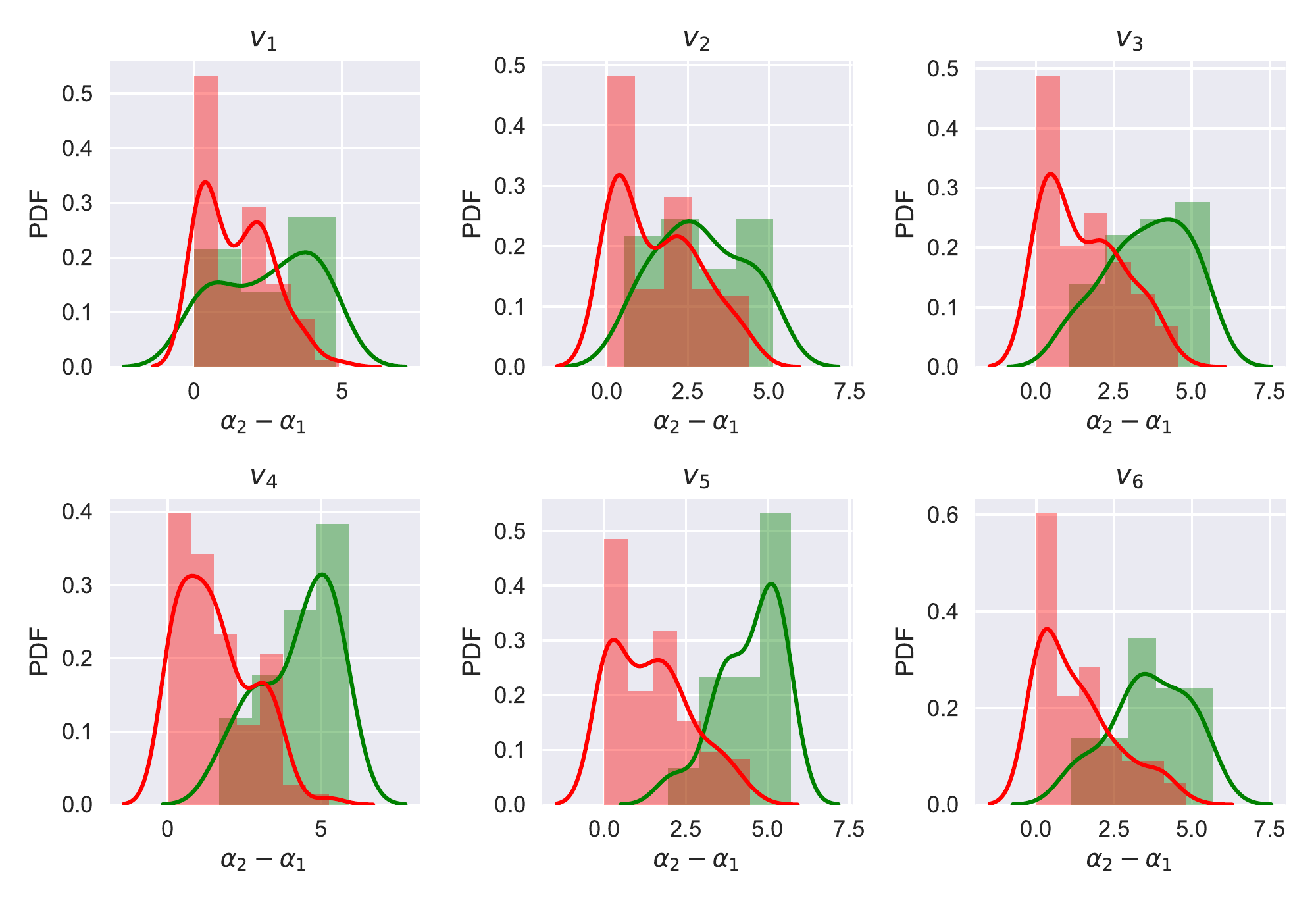}
\caption{\label{alpha_diff_hist}  Distributions of the difference $\alpha_2-\alpha_1$ for healthy(green) and unhealthy(red) groups for different channels. This difference is a measure of the complexity of dynamics underlying the ECG, as described in the text and thus, healthy hearts seem to be more complex than the unhealthy ones.}
\end{center}
\end{figure}

The difference $\alpha_2-\alpha_1$, that measures the width of the $f(\alpha)$ spectrum provides the range of scaling indices. In Fig.~\ref{alpha_diff_hist}, we show the distributions of this difference for healthy and unhealthy groups. It can be concluded that the complexity tends to be more for healthy hearts across all the channels.

\subsection{Parameter planes}
\begin{figure}[h]
\includegraphics[width=1\columnwidth]{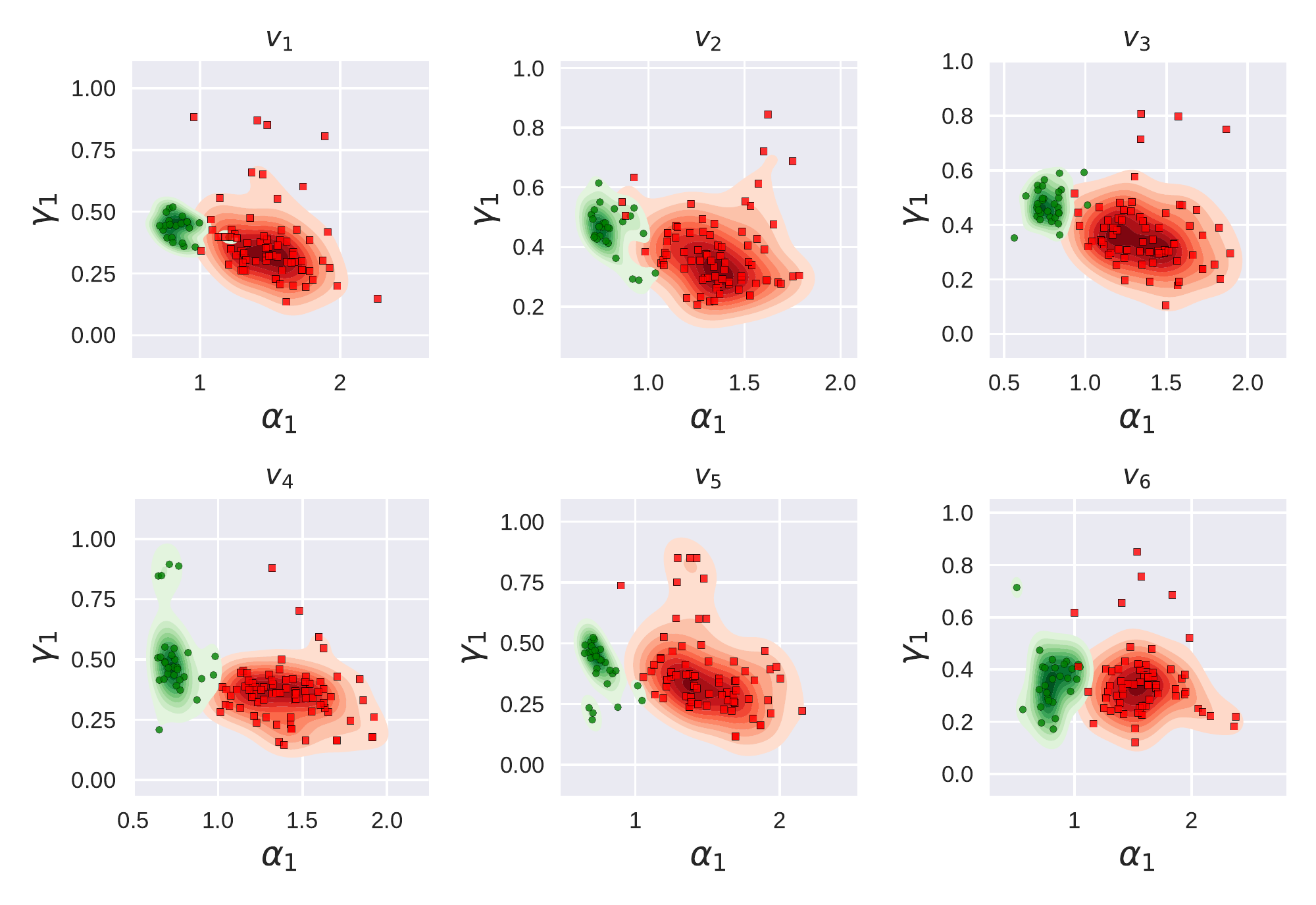}
\caption{\label{alpha_gamma}  Scatter plots for electrodes $v_1$ to $v_6$ in $\alpha_1$-$\gamma_1$ parameter planes. Green circles in the figure represent the healthy subjects and red squares represent patients. Overall, healthy cases are seen to be clustered whereas the patients are separated from healthy ones and are scattered.}
\end{figure}

In Fig.~\ref{alpha_gamma}, we present the scatter plots of $\alpha_1$ and $\gamma_1$ values for electrodes $v_1$ to $v_6$. The green circles in these plots represent the subjects identified as healthy in the PhysioNet database and the red squares represent the unhealthy ones. A few cases, for which the good fit could not be obtained have been discarded while plotting these parameter planes. For the purpose of visualization, we also show estimated kernel densities for the two groups as a background. As can be seen from these plots, multifractal analysis seems to have picked up almost every case correctly by separating the healthy and unhealthy cases into different clusters in $\alpha_1$-$\gamma_1$ planes. 

\subsection{Blind data testing and success rate}
\begin{figure}[h]
\begin{center}
\includegraphics[width=\columnwidth]{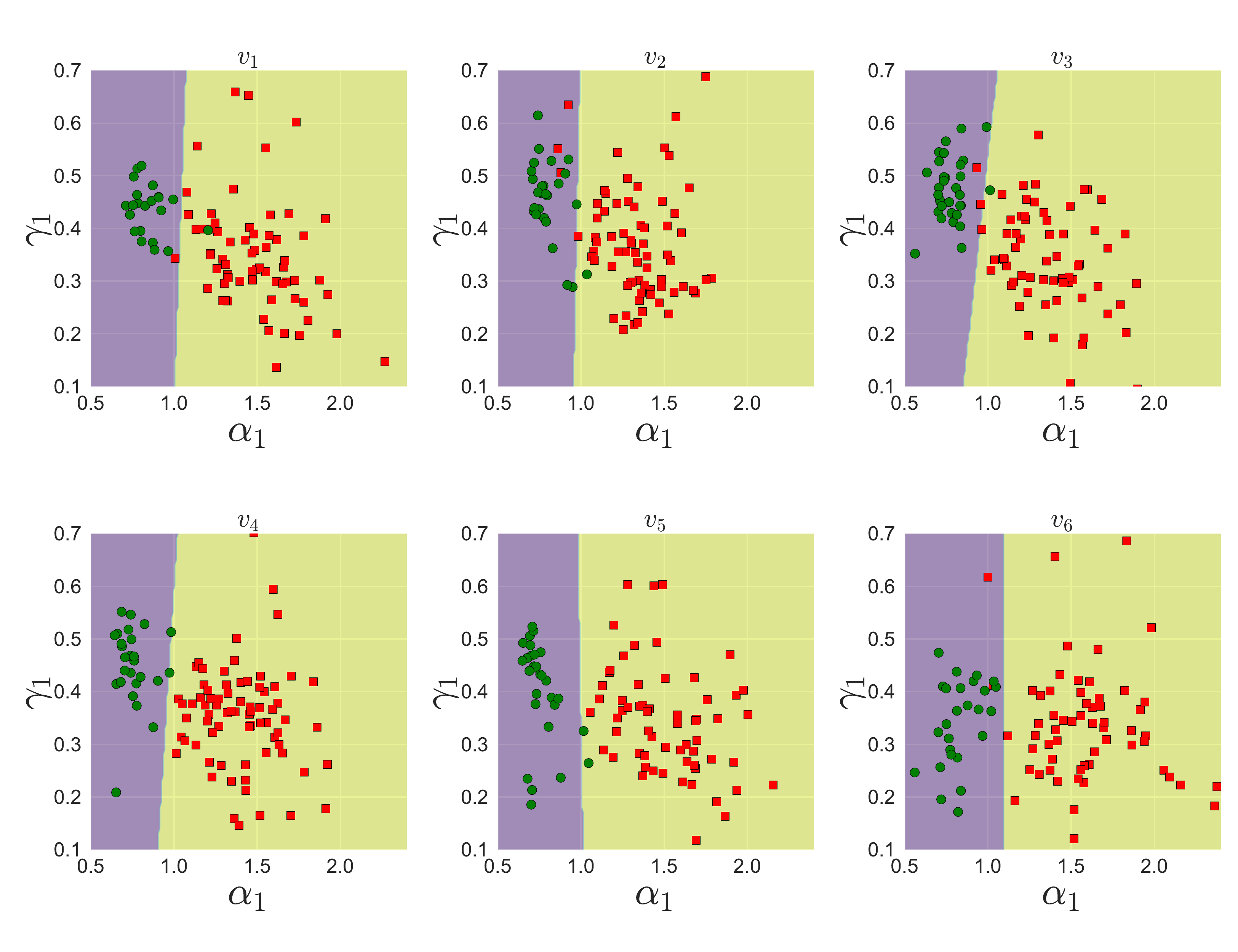}
\caption{\label{svc_rbf} Regions of the parameter plane $\alpha_1$-$\gamma_1$ that separate healthy and unhealthy regions obtained using SVC algorithm for electrodes $v_1$ to $v_6$. The actual data points are also shown for comparison with the regions.}
\end{center}
\end{figure}

The separation of the two groups into two different clusters becomes useful only if it helps us to predict the group label (healthy or unhealthy) for a new unseen data. This is a standard problem in the theory of machine learning and we use a particular algorithm called a ``support vector clustering'' or SVC using RBF kernel \cite{bishop2006pattern} to find out the regions in $\alpha_1$-$\gamma_1$ planes corresponding to the two groups. The known group labels are used as a training data for the algorithm to identify healthy and unhealthy cases and then the algorithm is asked to divide the parameter plane into two regions. The regions so obtained for different channels are shown in Fig.~\ref{svc_rbf}.

\begin{figure}[h]
\begin{center}
\includegraphics[width=\columnwidth]{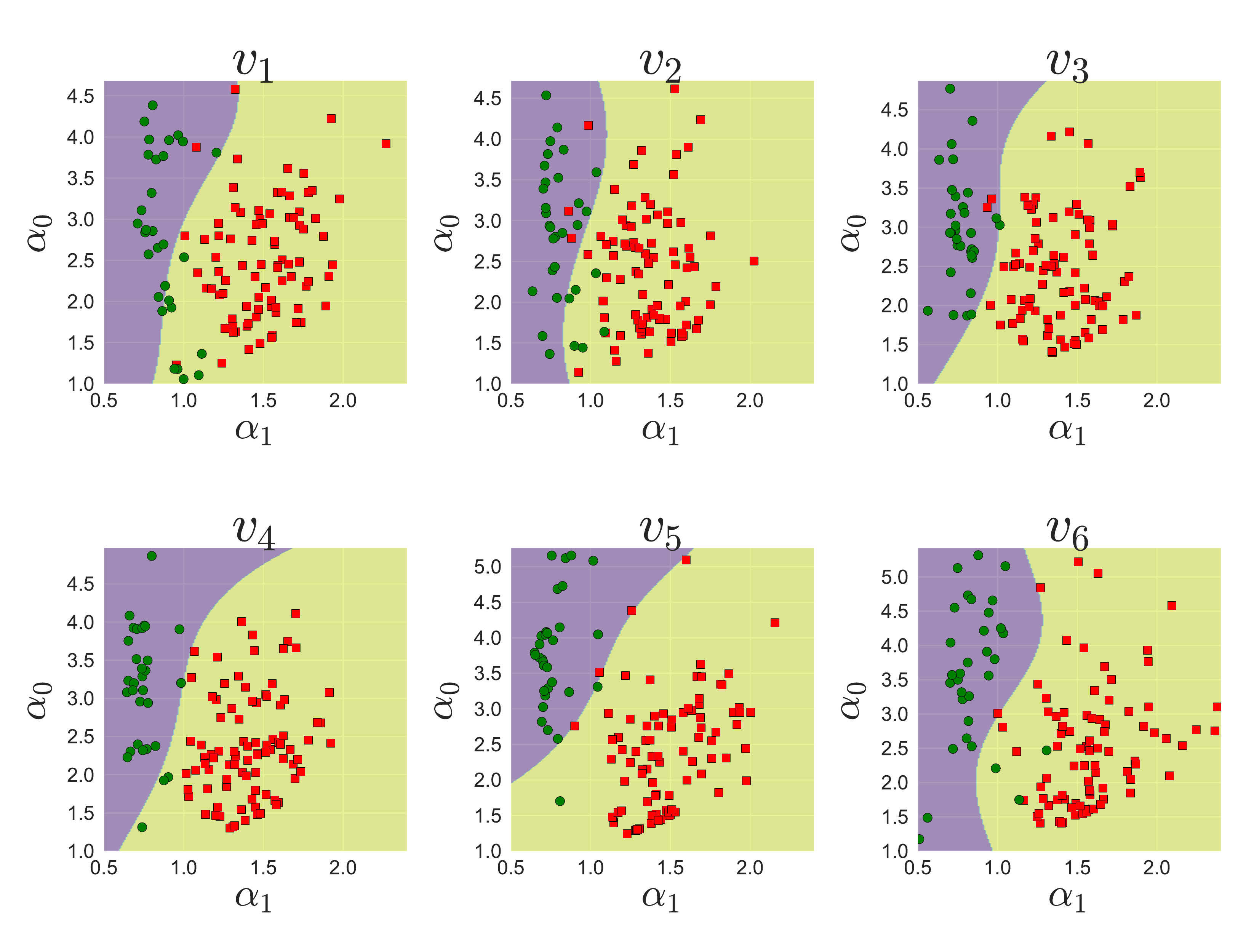}
\caption{\label{svc_alpha} Regions corresponding to healthy and unhealthy cases in $\alpha_1$-$\alpha_0$ plane found using SVC algorithm.}
\end{center}
\end{figure}

As another set of quantifiers, we now consider the parameter plane $\alpha_1$-$\alpha_0$, where $\alpha_0$ is the $\alpha$ value corresponding to the maximum value of $f(\alpha)$ curve and perform a similar analysis. The resulting regions are shown in Fig.~\ref{svc_alpha}.

The regions shown in Fig.~\ref{svc_rbf} and Fig.~\ref{svc_alpha} are obtained by training the SVC algorithm by using the whole data. However, this doesn't tell us how well the algorithm would perform when given an unseen data. To check this, we split the whole data into two parts: training set and test set. We then train the algorithm on the training data and then ask it to predict the labels for the test data. To measure the success we calculate true positive rate ($t_p$) and true negative rate ($t_n$). True positive rate in this case is defined as the fraction of correctly identified healthy cases out of the actual healthy cases. Similarly, true negative rate is defined as the fraction of the correctly identified unhealthy cases out of the actual unhealthy cases. We then define the accuracy of the algorithm to be: $\text{Accuracy} = t_p \times t_n$.

\begin{figure}[h]
\begin{center}
\includegraphics[trim=50 0 20 0, clip = true, width=\columnwidth]{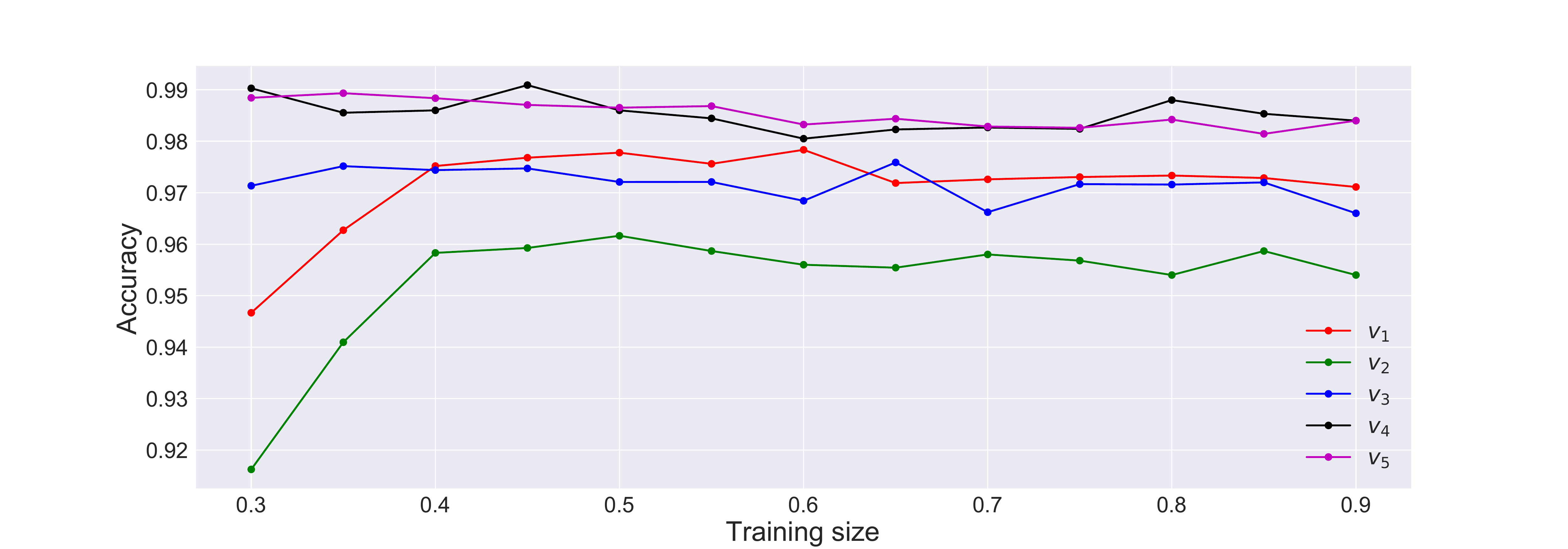}
\caption{\label{svc_scores}  Accuracy of the SVC algorithm as a function of fraction of training data set. It can be seen that even for the low sizes of the training sample, the algorithm predicts the class labels with extremely high accuracy ($>90\%$).}
\end{center}
\end{figure}

This definition of the accuracy makes sure that both cases are predicted reasonably accurately since even if one of them is low the accuracy becomes low. In particular, if the algorithm labels all cases to be of the single group, the accuracy becomes zero. The value of the accuracy also depends on how the data is split and so, we average the value over ten random realizations of the splitting. These average accuracies, calculated for $\alpha_1$-$\gamma_1$ planes, as a function of size of the training set for different channels are shown in Fig.~\ref{svc_scores}. It can be seen that even for low amount of training data, the accuracy is quite high. 

\section{Comparison with beat replicated data}
The results described in the previous section show that the multifractal analysis is extremely successful for separating healthy and unhealthy classes. Thus, given an ECG time series of a person that is not known to be in one or the other group a priori, we can calculate the corresponding $\alpha_1$-$\gamma_1$ values and then using its location in this parameter plane, we can predict the class the person belongs to with sufficient confidence. However, in general, definition of healthy and unhealthy can be quite subjective. For example, in the healthy class, ECG characteristics in general differ based on age, gender, habits, life styles etc. This makes comparing the measures of one set with that of another somewhat arbitrary unless both are with the same age, gender, habits etc, which is not always very practical. Therefore we introduce a finer method of analysis, by checking how the multifractal properties of a given data compare with that of the beat replicated data generated from a single beat in the same ECG. This would in a way compare each data within itself and the range of variations can be used effectively as a quantifier for normal and abnormal hearts.

\begin{figure}[h]
\begin{center}
\includegraphics[width=0.9\columnwidth]{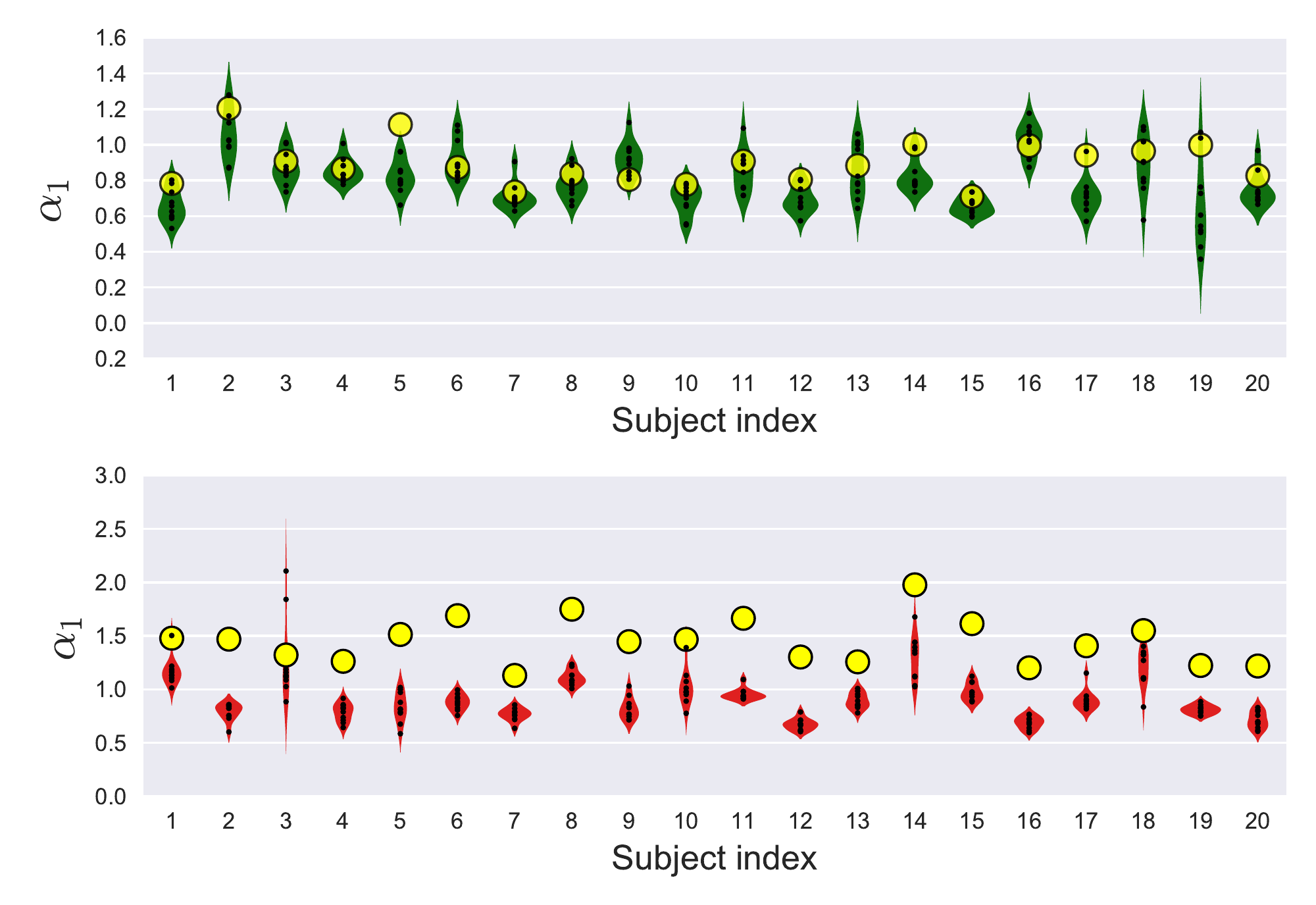}
\caption{\label{replicate_violin} Violin-plot of $\alpha_1$ values calculated by replicating several randomly chosen beats for few healthy (top panel) and unhealthy (bottom panel) subjects. Each label on $x$-axis represents a subject and the $\alpha_1$ values for time series obtained by replicating different beats of that person are plotted along $y$-axis. The bigger yellow markers in the plots represent the $\alpha_1$ values obtained from the complete time series. It can be seen that the actual $\alpha_1$ values for patients tend to fall outside the replicated values.}
\end{center}
\end{figure}

\begin{figure}[h]
\begin{center}
\includegraphics[width=\columnwidth]{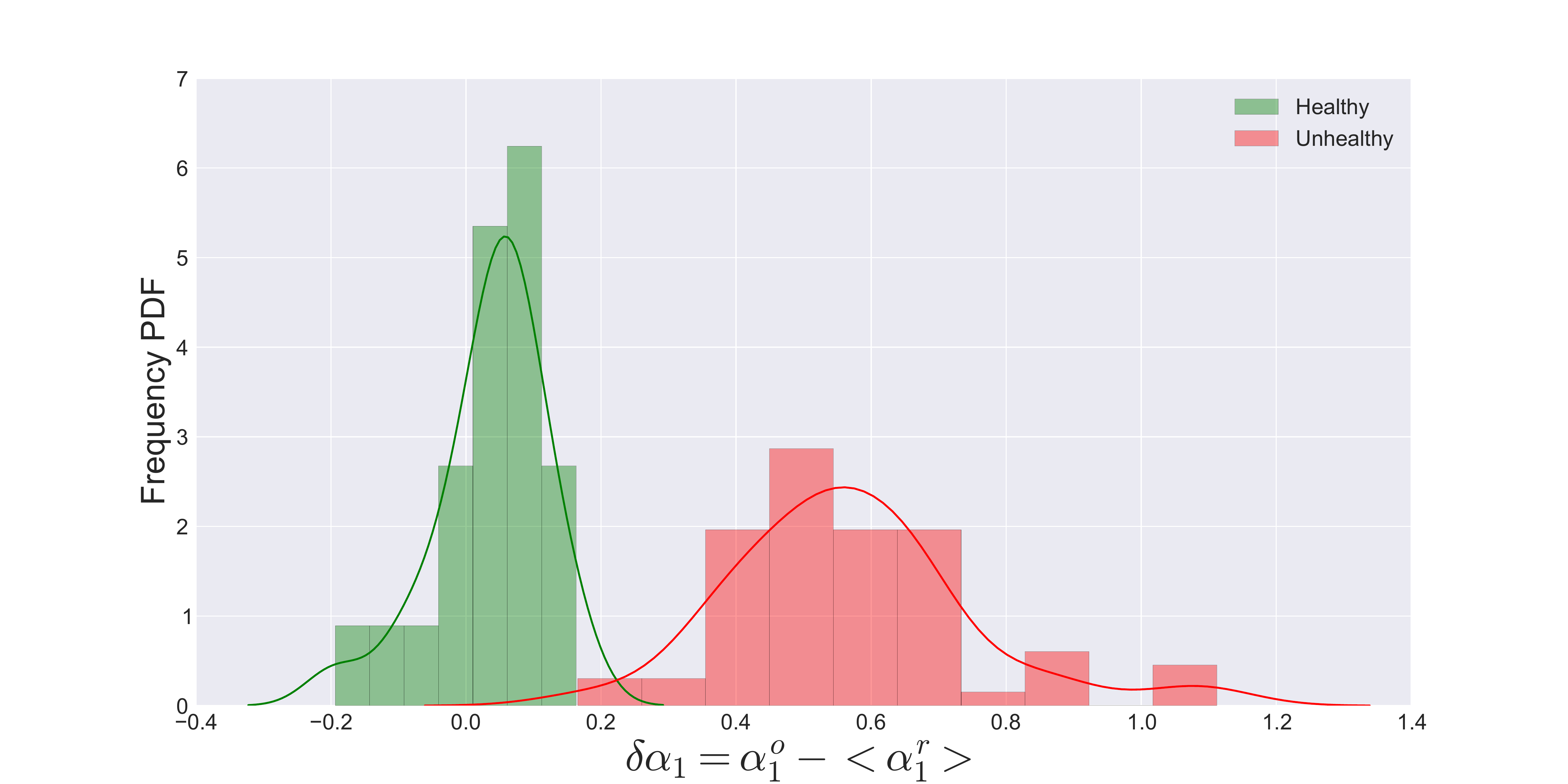}
\caption{\label{diff_alpha1}  Distributions of the difference between the mean of the replicated $\alpha_1$ values and the $\alpha_1$ value for the full time series for healthy and unhealthy groups.}
\end{center}
\end{figure}

We extract $10$ different randomly chosen beats from each signal (See Methods). Then we replicate each of these beats to get time series of approximately the same sizes as that of the original time series and perform multifractal analysis for each one of these to obtain the parameters $\alpha$ and $\gamma$. In Fig.~\ref{replicate_violin}, we show distributions of $\alpha_1$ values for $20$ randomly chosen subjects from each group. For comparison, we also show the original $\alpha_1$ value in each case (yellow circles). As can be seen, the actual $\alpha_1$ values in case of healthy subjects tend to coincide with the $\alpha_1$ values for beat replicated time series. On the other hand, the actual $\alpha_1$ values in case of unhealthy cases tend to be quite far from the mean of the replicated $\alpha_1$ values. To make this quantitative, we define $\delta\alpha_1$ to be the difference between the mean of the replicated $\alpha_1$ values and the $\alpha_1$ value for the full time series. The histograms of this quantity for two groups are shown in Fig.~\ref{diff_alpha1}. It can be seen that the distributions are quite separated from each other. This implies that one can distinguish the two classes solely on the basis of the comparison of values for the data with its own beat replicated data sets. 

\matmethods{

\subsection{Data acquisition}
For our analysis presented here, we have used Data from ``PhysioNet'' Resource with its PhysioBank archive \cite{goldberger2000physiobank}. In total, we included data for $97$ abnormal subjects and $32$ healthy subjects. Each dataset consists of $15$ channels which correspond to different electrodes, the conventional $12$ leads ($i, ii, iii, avr, avl, avf, v_1, v_2, v_3, v_4, v_5, v_6$) together with the $3$ Frank leads ($v_x, v_y, v_z$). Out of these, in this work we concentrate only on six of the channels $v_1$ to $v_6$ that correspond to electrodes which are placed directly on the chest. Each signal corresponds to a real time of $60$ seconds and is digitized at $1000$ samples per second to obtain in total $60000$ points per signal.  

Among the $97$ patient data available to us, $79$ suffer from Myocardial Infraction (MI), $6$ suffer from Cardiomyopathy, $4$ suffer from Myocarditis, $2$ suffer from Dysrhythmia and $1$ from Hypertrophy while for the remaining $5$, the disease information is not available. The subjects in the healthy class have age values distributed between $24$ and $69$ whereas the age values in the unhealthy class are distributed between $41$ and $86$. 

\subsection{Detrending of the signals}
The ECG signals often contain global trends as shown for a typical data in the top panel of Fig.~\ref{ecg_detrended_ts}. As part of the pre processing, we first remove these trends as described below and the de-trended data thus obtained after removing the global trends is shown in Fig.~\ref{ecg_detrended_ts}.

To remove the undesirable trends, we fit a polynomial of a certain degree to the signal, which is then subtracted from the actual signal to get the de-trended signal. To choose the appropriate value of the degree $n$ to be used for the fitting polynomial, we define a deviation $\delta_{(n)}$ of the original signal from the detrended signal as:

\begin{equation}
\delta_{(n)} = \frac{1}{N}\sum\limits_{i=1}^{N}(x_o(i)-x_{(n)}(i))^2
\end{equation}

\begin{figure}
\begin{center}
\includegraphics[width=0.8\columnwidth]{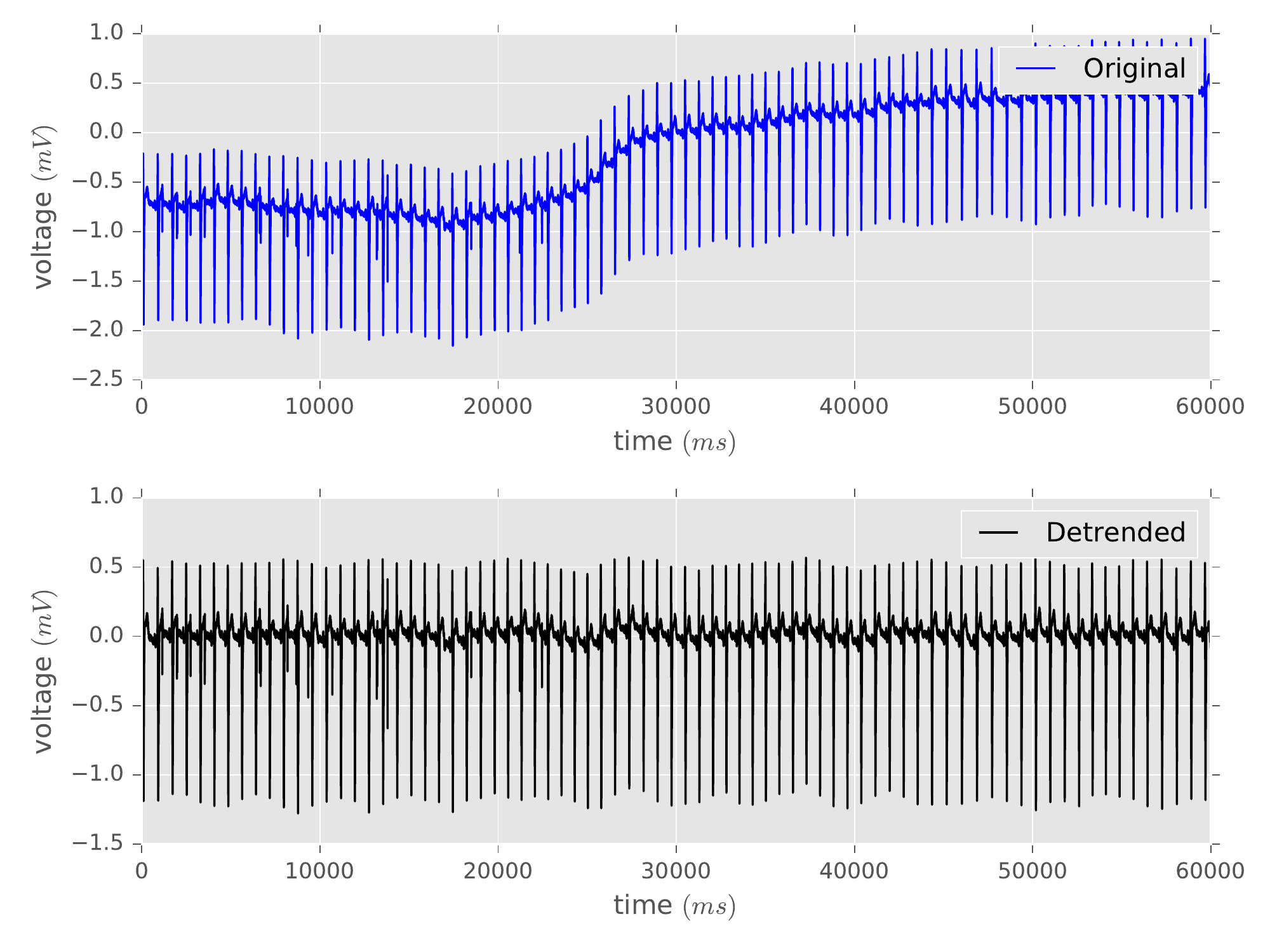}
\caption{\label{ecg_detrended_ts} Time series of a randomly chosen subject before detrending (top panel) and after detrending (bottom panel). As explained in the text, the global trend is removed by fitting a polynomial to the original data. }
\end{center}
\end{figure}
We find that $\delta_{(n)}$ saturates as we vary $n$ and hence we can choose $n$ after the saturation point for a given signal. Based on this, for all the datasets we use $n=20$ to detrend them. 


\subsection*{Embedding and phase space reconstruction}
For uniformity, all the values in the time series $c(t_k)$ are first scaled between $0$ and $1$ by using a transformation of ``compression'':
\begin{equation}
s(t_k) = \frac{c(t_k)-c_{\text{min}}}{c_{\text{max}}-c_{\text{min}}}
\end{equation}
where $c_{\text{min}}$ and $c_{\text{max}}$ are minimum and maximum values in the time series $c(t_k)$ respectively. 
Each time series $s(t_k)$ is then embedded into an $M$ dimensional space, by constructing vectors as:
\begin{equation}
\vec{x}_i = [s(t_i), s(t_i + \tau), s(t_i + 2\tau), \cdots, s(t_i + (M-1)\tau)]
\end{equation}
Here a time delay $\tau$ is the time, measured in units of sampling rate $\triangle = t_{i+1}-t_i$, at which autocorrelation of the signal falls to $1/e$ of its original value \cite{harikrishnan2006non}. It is easy to see that there are in total $N-(M-1)\tau$ embedded vectors. Taken's embedding theorem dictates that the phase space trajectories or attractor obtained from these vectors have the same topological properties as that of the original system\cite{takens1981dynamical}.

\subsection{Extracting a single beat from an ECG time series}
Identifying a single beat in an ECG signal is tricky since a beat cannot be defined as a pattern that repeats with exact periodicity in the ECG signal. However, it is easy to see that ECG signals do have a certain approximate periodicity because of the presence of beats. For the data used, in units of milliseconds, the individual beats repeat with a period $T\in (600, 1500)$. We then calculate the autocorrelation of the time series and find out the highest peak in this range. The corresponding time value is then taken as the period of the signal and the same is used to extract a single beat from the time series. 

}

\section{Discussion}
We report the results of a detailed multifractal analysis of discretized ECG data for two sets of healthy and unhealthy subjects. Our study establishes the highly complex fractal nature of a healthy heart, which gets reduced due to any abnormality in its functions.  We could show that the measures derived from the multifractal spectrum can detect abnormalities in heart dynamics with a reasonable level of accuracy. The fact that this is achieved with short time ECG recordings of a significantly small amount of time enhances the scope for its applicability. Moreover the analysis is totally objective and the results are quantitative.

The novelty of our approach is that the subtle variations in the shapes of the beats are captured into readable quantities or values which will be much more reliable than conclusions derived from visual inspections. This information can be used to assess the status of health and risk level of the heart easily. The fluctuations of the beat-to-beat dynamics as revealed in the ECG exhibit a rich complexity that, as carefully characterized in this work, can give indications of cardiac malfunctions. Our analysis, using beat replicated data, helps in comparing quantifiers of one ECG with variations within itself. The study is thus an advancement both in the basic understanding of the fractal nature underlying a complex system like heart and its abnormal states as well as in designing clinically useful practical tools in effective diagnostics and therapy. 

\showmatmethods{} 

\acknow{The authors acknowledge the financial support from Dept. of Sci. and Tech., Govt. of India, through a Research Grant No.
EMR/2014/000876. We also acknowledge the Physiobank data base (www.physionet.org/physiobank/database/) for the ECG data used in the study reported here.}

\showacknow{} 



\end{document}